\documentclass{aa}
\usepackage{txfonts}
\usepackage{natbib}
\usepackage{graphicx}
\bibpunct{(}{)}{;}{a}{}{,} % to follow the A&A style
\usepackage{longtable}

\begin{document}

\title{Transneptunian objects and Centaurs from light curves}

%\thanks{Based on observations collected at the Centro
%Astron\'{o}mico Hispano Alem\'{a}n (CAHA) at Calar Alto, operated
%jointly by the Max-Planck Institut f\"{u}r Astronomie and the
%Instituto de Astrof\'{\i}sica de Andaluc\'{\i}a (CSIC) and the
%Sierra Nevada Observatory, managed by the Instituto de
%Astrof\'{\i}sica de Andaluc\'{\i}a (CSIC).}}

\author{Duffard, R. \inst{1}
   \and Ortiz, J.L. \inst{1}
   \and Thirouin, A. \inst{1} \and Santos-Sanz, P.\inst{1} \and N.
   Morales \inst{1}}
 \offprints{Duffard, R. \email{duffard@iaa.es}}
\institute{$^1$ Instituto de Astrof\'{\i}sica  de Andaluc\'{\i}a -
CSIC, Apt 3004, 18080  Granada,  Spain }

\titlerunning{KBOs CCD photometry}

\date{received/accepted}

\abstract{}{We compile and analyze an extended database of light curve
parameters scattered in the literature to search for correlations and study
physical properties, including internal structure constraints.} {We analyze
a vast light curve database by obtaining mean rotational properties of the
entire sample, determining the spin frequency distribution and comparing
those data with a simple model based on hydrostatic equilibrium.} {For the
rotation periods, the mean value obtained is 6.95 h for the whole sample,
6.88 h for the Trans-neptunian objects (TNOs) alone and 6.75 h for the
Centaurs. From Maxwellian fits to the rotational frequencies distribution
the mean rotation rates are 7.35 h for the entire sample, 7.71 h for the
TNOs alone and 8.95 h for the Centaurs. These results are obtained by taking
into account the criteria of considering a single-peak light curve for
objects with amplitudes lower than 0.15 mag and a double-peak light curve
for objects with variability $>$0.15mag. We investigate the effect of using
different values other than 0.15mag for the transition threshold from
albedo-caused light curves to shape-caused light curves. The best Maxwellian
fits were obtained with the threshold between 0.10 and 0.15 mag. The mean
light-curve amplitude for the entire sample is 0.26 mag, 0.25 mag for TNOs
only, and 0.26 mag for the Centaurs. The Period versus B-V color shows a
correlation that suggests that objects with shorter rotation periods may
have suffered more collisions than objects with larger ones. The amplitude
versus H$_v{}$ correlation clearly indicates that the smaller (and
collisionally evolved) objects are more elongated than the bigger ones.}
{From the model results, it appears that hydrostatic equilibrium can explain
the statistical results of almost the entire sample, which means hydrostatic
equilibrium is probably reached by almost all TNOs in the H range [-1,7].
This implies that for plausible albedos of 0.04 to 0.20, objects with
diameters from 300 km to even 100 km would likely be in equilibrium. Thus,
the great majority of objects would qualify as being dwarf planets because
they would meet the hydrostatic equilibrium condition. The best model
density corresponds to 1100 kg/m$^3$.}

 \maketitle

%
%________________________________________________________________

\section{Introduction}

The belt of remnant planetesimals in the so-called Kuiper belt, whose
existence was confirmed observationally in 1992 \citep{Jewitt1993}, is an
important source of knowledge about the formation and early evolution of the
Solar System. Features in the orbital distribution of the belt can provide
important information about the early dynamical processes such as planetary
migration, which was merely untested theories around 25 years ago
\citep{Fernandez1984} that now can be regarded as being well proven. The
detailed architecture of the Kuiper belt might also be compatible with a
very unstable phase when Jupiter and Saturn entered into resonance
\citep{Tsiganis2005, Morbidelli2005, Gomes2005} around 800Myr after the
Solar System formation. Although this particular dynamical scenario is still
disputed, there is no question that the architecture of the transneptunian
belt can offer invaluable clues about the dynamical mechanisms acting in the
early Solar System. In addition to dynamical inferences from the orbital
characteristics of the TNOs, much can be learnt about the physical processes
that took place in the solar nebula by means of the study of the physical
properties of TNOs and related groups of objects. Since the TNOs are
understood to be mixtures of ices and rocks from which the Centaurs and
ultimately the Jupiter family comets (JFCs) originate \citep{Jewitt2004b},
ground and space-based observations of all these populations can give us
clues about the physics of the outer Solar System. From the ``Nice"
dynamical model \citep{Tsiganis2005, Morbidelli2005, Gomes2005}, it is also
likely that the Jupiter trojans are closely related to the TNOs
\citep{Morbidelli2005}, but this is not yet well established.

These objects are regarded as the least evolved objects in the Solar System,
but they have experienced many different types of physical processes that
have altered them considerably. These processes range from collisions, to
space weathering. The collisional processes have probably left clear
imprints in both the spin distribution and in the light-curve amplitude
distribution of TNOs and Centaurs. The currently accepted theory based on
the study of \cite{DavisFarinella1997} on collisional evolution of the TNOs
is that objects of sizes larger than 100 km are very slightly collisionally
evolved, whereas the smaller fragments have experienced more collisional
events and hence should have their rotations and shapes more significantly
altered. This picture is also consistent with newer collisional evolution
models \citep{Benavidez2009}. However, detailed modelling of the evolution
of the spin rates related to collisional models has not yet been carried out
for the transneptunian belt.

Due to the dedicated efforts of two main groups studying short-term
variability \citep{Gutierrez2001, Ortiz2003a, Ortiz2003b, Ortiz2004,
Ortiz2006, Ortiz2007, Moullet2008, Duffard2008,Sheppard2007,
Sheppard-Jewitt2002, Sheppard2000} and occasional contributions by other
authors, a database of about 40 rotation periods and 80 light-curve
amplitudes could be extracted from the literature in 2008. Thus, we now have
sufficient numbers of TNOs light curves to start analyzing them from a
statistical point of view. In a review by \cite{Sheppard2008}, those data
were compiled and some conclusions were outlined. After that compilation, a
new set of light curves from 29 TNOs and Centaurs became available
\citep{Thirouin2009} and those authors also presented evidence of possible
biases that can be present in the light curve database gathered from the
literature.

We also developed a simple model to interpret rotation rates and light-curve
amplitudes together, to obtain information about densities and tensile
strengths. Instead of analyzing each body separately and trying to retrieve
information about its density and cohesion by using its spin rate and
light-curve amplitude (and an assumed viewing angle), we address the mean
internal properties by analyzing the population as a whole. In other words,
we developed a model that can simultaneously interpret the distribution of
spin rates and the distribution of light-curve amplitudes.

In this paper, light curve parameters of a vast sample are compiled and
analyzed. We focus on the analysis of the period and shape distribution
presented in Sect. 2. Then, with the current data, we complete a correlation
analysis with several orbital parameters in Sect. 3. Internal properties of
the TNOs are derived by applying a simple model in Sect. 4, a discussion of
the results is presented in Sect. 5. Finally, the conclusions are presented
in Sect. 6.

%
%________________________________________________________________

\section{Period and light-curve amplitude distributions}

One important current limitation is that the light curves of this kind of
objects are accessible only with medium-size (2 meters) or larger
telescopes. The telescope time required for this type of program is
difficult to obtain, and the observed objects are the brightest and largest
ones. Only several light curves were observed with the Hubble Space
Telescope (HST) reaching a lower limit of 20-100 km in diameter
\citep{Trilling-Bernstein2006}. The current Kuiper belt object (KBO)
absolute magnitude for which a good quality light curve can be derived (with
a 2 meter telescope), is H $<$ 9 for TNOs and H$<$ 11 for Centaurs. Smaller
KBOs can only be observed with space telescopes.

Using the literature and our recent results \citep{Thirouin2009}, it is
possible to create a database of light curves with rotational periods and
peak-to-peak amplitudes. The special case of Pluto, whose spin rate has been
altered by tidal dissipation caused by Charon, has been removed from the
sample. A compilation of 102 objects, updated on March 2009, is presented in
Table \ref{table1}. From that list, only 72 objects have good enough period
and light-curve amplitude estimates.

When looking and analyzing the current TNO and Centaur light curves, the
first noticeable characteristic in most of the derived light curves is that
the amplitudes are very small ($<$0.15 mag). The criterium of placing a
limit of 0.15 mag was introduced by \cite{Sheppard-Jewitt2002} and used by
other authors \citep{Lacerda-Luu2006,Ortiz2003a,Ortiz2003b} to distinguish
the ``photometrically flat" curves. The brightness changes are probably
caused by the combined effect of rotation, irregular shape and/or albedo
markings on the surface. Thus, the abundance of low amplitude light curves
is indicative of a large abundance of quasi-spherical bodies with
homogeneous surfaces. However, this is to some degree a selection effect
because, as mentioned above, the true sample is dominated by ``bright"
objects, which are presumably the largest ones and nearly spherical.

\begin{figure}
\resizebox{18cm}{!}{\includegraphics[angle=0]{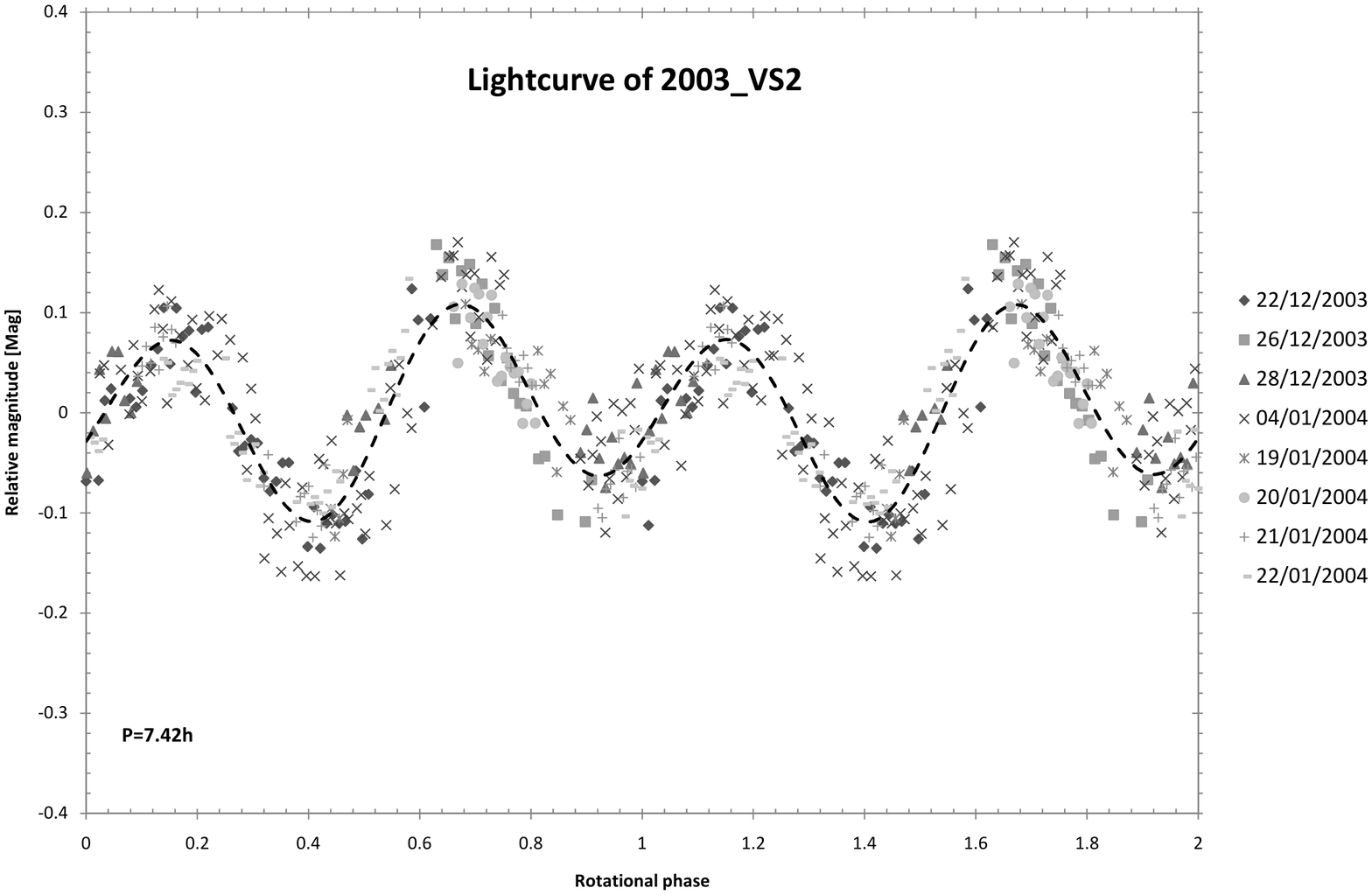}}
\caption{Example of a KBO light curve from \cite{Thirouin2009} obtained
during several nights.}
         \label{fig01}
   \end{figure}

An example of the kind of light curve studied here is presented in Fig.
\ref{fig01}. In this example, data of the object 2003 VS$_{2}$ over several
nights was observed and the composite light curve was compiled to determine
the rotational period and amplitude. Unfortunately, it is not always
possible to obtain a well determined period and/or amplitude. Authors
sometimes place a constraint on the values of periods and/or amplitudes, and
we note that the literature is populated by these kinds of results. In Table
\ref{table1}, we list all objects with published rotational period and/or
light curve amplitude or constraints in one of these values. In any case,
for all the considerations in the paper only the objects with a well
determined period/amplitude pair are taken into account. That is the case
for 72 objects. In case of multiple determinations of the period and/or
amplitude, the criteria selected the preferred value by the author who
published that value. If no preferred value is mentioned, the mean of the
listed values was used.

\subsection{Amplitude distribution}

The KBOs albedo may not be uniform across their surfaces, causing the
apparent magnitude to vary as the different albedo markings on the KBOs
surface rotate in and out of our line of sight. Albedo or surface variations
of the object are usually responsible for less than a 20\% difference from
maximum to minimum brightness of an object \citep{Degewij1979,
Magnusson1991} in the asteroid case. In \cite{Thirouin2009}, it is estimated
that the typical TNO variability caused by albedo markings is around 0.1
mag. The reasons for the apparently small amount of variability due to
albedo markings on the TNO surfaces compared to that of the asteroids may be
related to the icier surfaces of the TNOs than those of the asteroids.
However, also the processing of the surfaces is different, because different
mechanisms can take place. Thus, it is  unsurprising that the surface
reflectance variability may be different for TNOs than for asteroids. The
majority of low-amplitude light curves of large objects are believed to be
caused by this kind of surface variations \citep{Sheppard2008}.

On the other hand, shape variations or elongation of an object will cause
the effective radius of an object to our line of sight to change as the KBO
rotates. Smaller KBOs are expected to be structurally elongated. To date,
few small (20 - 100 km) KBOs have been observed with rotational variations,
and they appear to have larger light curve amplitudes
\citep{Trilling-Bernstein2006}.

\begin{figure}
\resizebox{18cm}{!}{\includegraphics[angle=0]{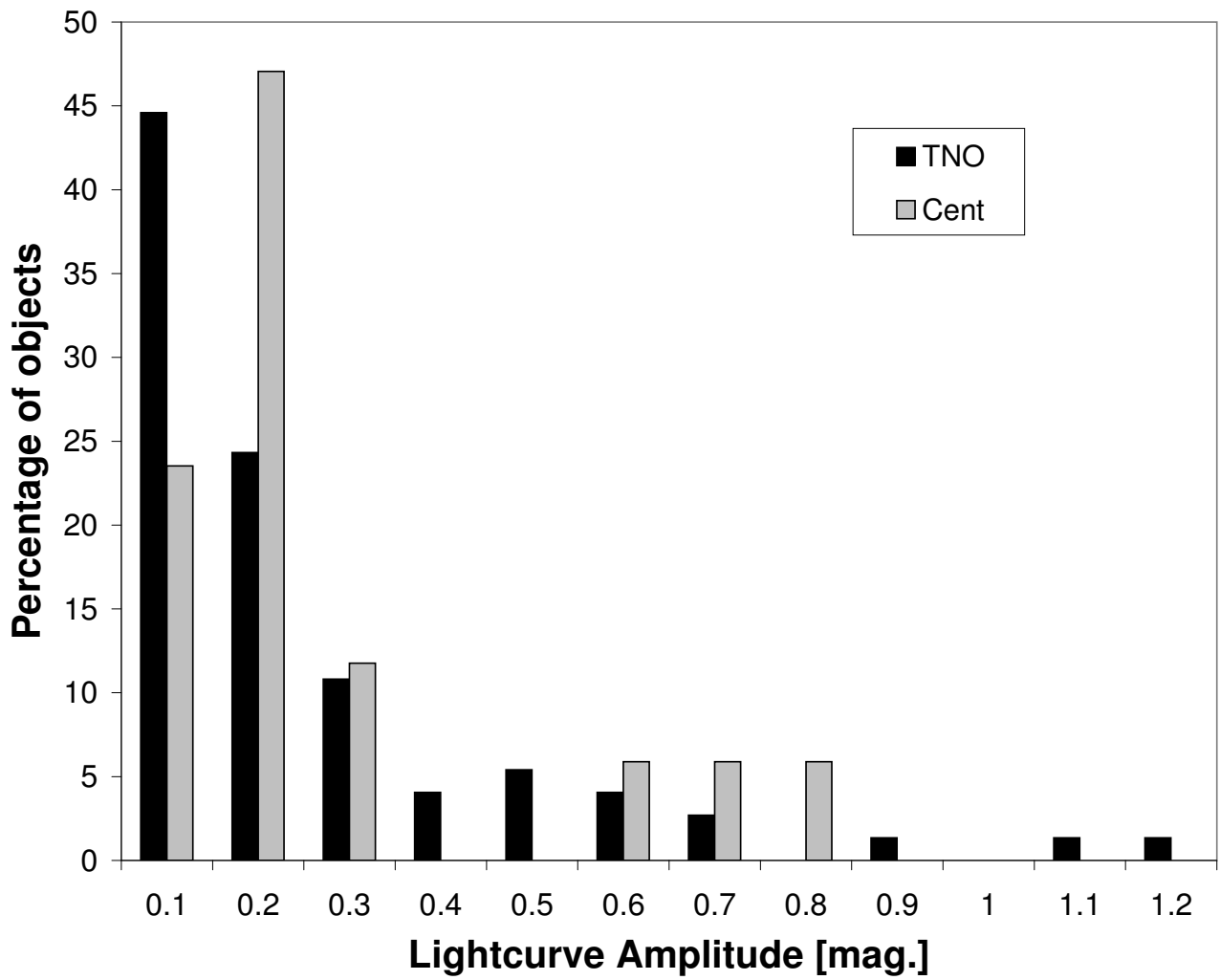}}
\caption{Percentage of objects vs. peak-to-peak light curve amplitude for
all TNOs (black) and Centaurs (grey).}
         \label{fig02}
   \end{figure}

In Fig. \ref{fig02}, we show the percentage of objects that have an
amplitude value within a specific peak-to-peak amplitude (in 0.1 mag bin) of
the well determined 91 amplitudes (74 for TNOs, and 17 for Centaurs) from
Table \ref{table1}. We note that some papers only report the amplitude, or
some restriction in the amplitude, but no period. That is the reason why the
number of amplitude determinations differ from the rotational period
determinations.

The first noticeable characteristic in that plot is that nearly 70\% of the
objects (in both cases, TNOs and Centaurs) have light-curve amplitudes
smaller than 0.2 mag. The reasons for observing a flat light curve could be
a spherical body with low contrast albedo marks on the surface, or a pole-on
configuration, that is, a rotational axis oriented toward the observer. This
last option is the least probable of all, so it is reasonable to say that
most of the observed TNOs are ``large" round objects with no significant
inhomogeneities on their surfaces. Objects with flat light curves correspond
in general to a less collisionally evolved population. Of course, the large
abundance of the low light-curve amplitude population could be to some
degree an observational bias because the KBOs with known rotational
properties are mainly the brightest ones. For the Centaurs, the vast
majority of objects have light-curve amplitudes between 0.1mag and 0.2mag,
which is indicative of some elongation in the shape of the bodies.

\subsection{Period distribution}

In the following analysis, all objects with known rotational period are
considered. An exception was made for Pluto, which was excluded. This
exception is because the long rotational period (due to the tidal locking of
the system) cannot be representative of the rotational properties of the
sample.

\begin{figure}
\resizebox{18cm}{!}{\includegraphics[angle=0]{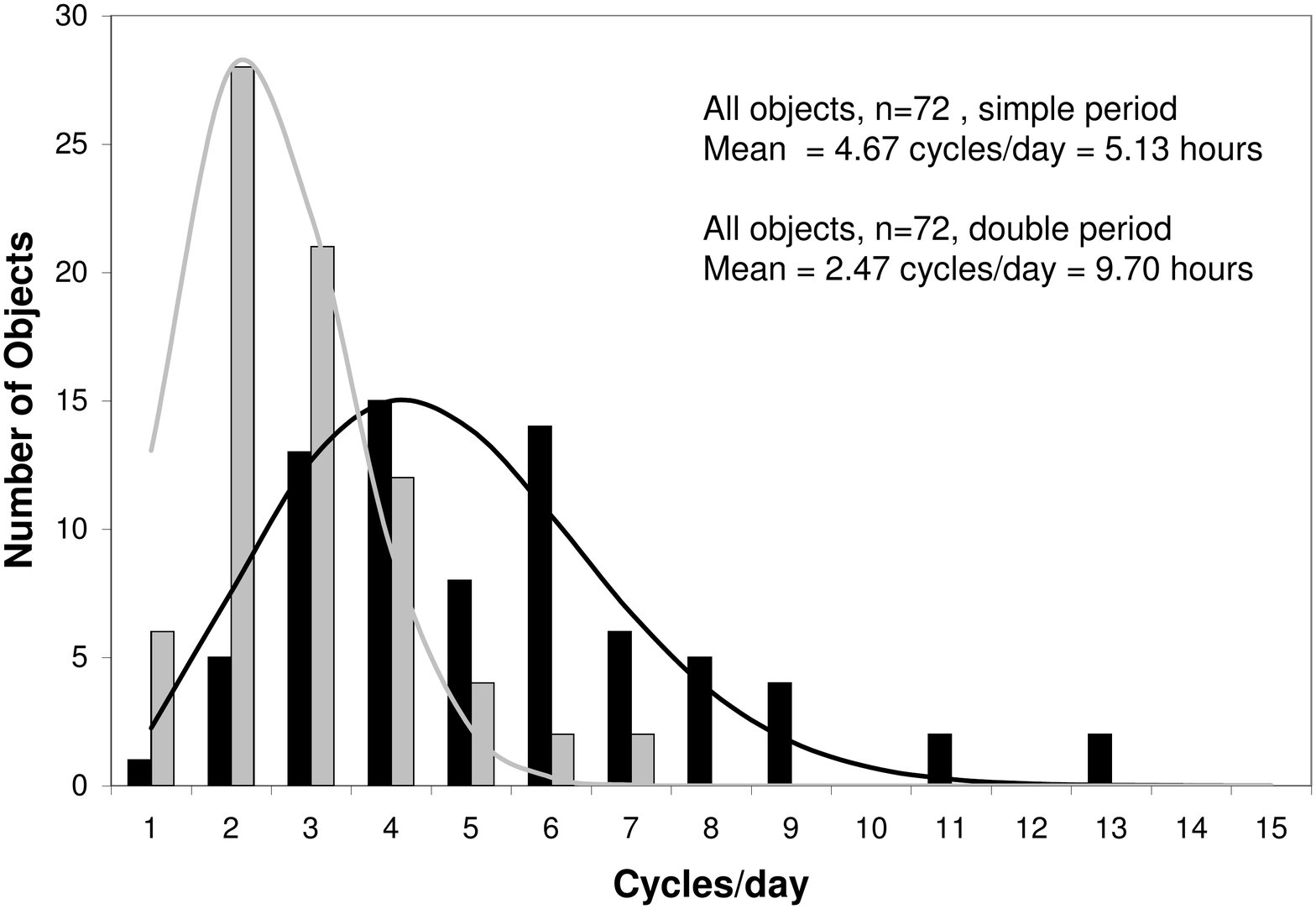}}
\caption{Histogram in rotational frequencies for all the objects considering
only a simple period rotation (black) and double period rotation (grey). A
maxwellian fit to the single peak period for all the object gives a mean
not-reliable lower period of 5.13 hours (black bars/line). Double peak
period is also represented (grey bars). The maxwellian fit to the data}
gives a mean not-reliable upper period of 9.70 hours.
         \label{fig03}
   \end{figure}

From KBO light curves alone it is difficult or even impossible to determine
whether the variability is caused by either albedo variations or an
elongated shape. Therefore, the rotation periods derived from the
photometric periods often have an uncertainty of a factor 2. A double-peaked
periodic light curve is expected to be seen in a non-spherical body, since
the projected cross-section should have two minima (short axis) and two
maxima (long axis) during one complete rotation of the KBO. Light curves
caused by surface inhomogeneities are single peaked. Figure \ref{fig03} (in
black) shows the histogram of rotational frequencies (single-peak rotational
period) for all the objects found in the literature. This histogram was
compiled by assuming that all objects have single-peaked light curves, which
is obviously not true and is an extreme case. For the well known cases in
which the light curves are certainly double-peaked, we used the rotation
period corresponding to half the true rotation period. Therefore, the
resulting distribution of rotation periods has a short mean value, which is
a crude lower limit to the mean rotation period. A Maxwellian curve was
fitted and overplotted in the histogram for a mean period of 5.13 hours.

The obvious next exercise is to plot the histogram of rotational frequencies
for all objects with double-peak light curves (in other words, we assume
that all rotation periods are twice the periods derived by a basic
periodogram analysis). This is again an unrealistic result, but provides an
upper limit to the rotational period distribution. This is the same as
considering all the light curves to be produced by non-spherical bodies. The
Maxwellian fit infers a mean frequency of 2.47 cycles/day or 9.70 hours
(data in grey).

\begin{figure}
\resizebox{18cm}{!}{\includegraphics[angle=0]{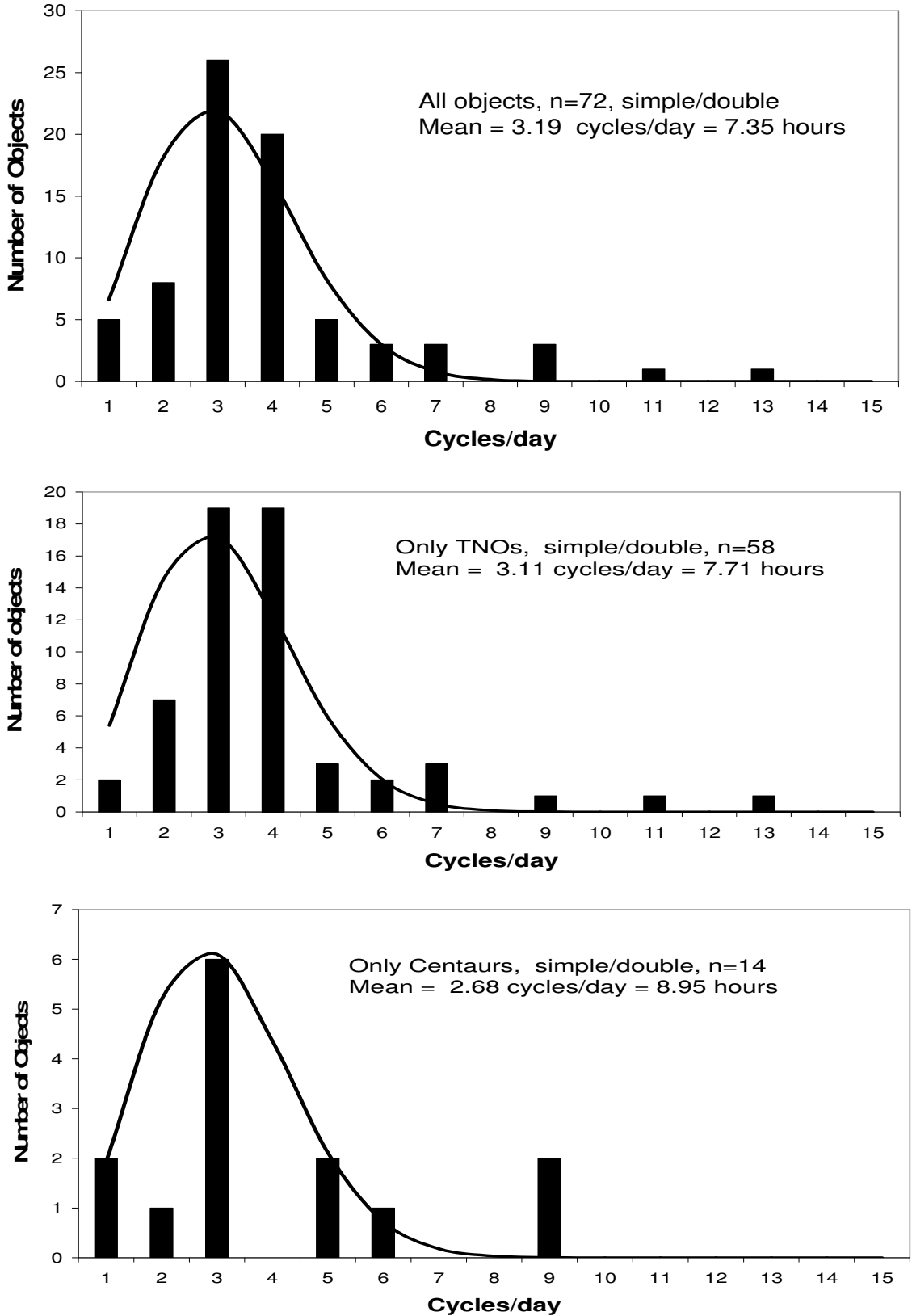}}
\caption{Histogram in cycles/day for the whole sample (top panel), the TNOs
(center panel) and Centaurs (lower panel). From maxwellian fits to the
rotational frequencies distribution the mean rotation rates are 7.35 hrs for
the entire sample, 7.71 hrs for the TNOs alone and 8.95 hrs for the
Centaurs. We chose to use the single peak period for the objects whose
amplitude is $<=$0.15~mag and double peak period for the objects whose
amplitude is $>$0.15~mag. }
         \label{fig04}
   \end{figure}

Finally, a more reliable histogram (presented in Fig. \ref{fig04}) was
compiled by assuming that all the light curves with amplitudes smaller than
or equal to 0.15mag are single peaked (this is equivalent to assuming that
the main cause for their rotational variation is albedo markings) and those
with amplitudes larger than 0.15mag are double peaked. Thus, with this
criterium we remove the factor of 2 ambiguity and derive a more realistic
distribution. There is also the possibility of deciding on a case-by-case
basis, but some cases are ambiguous.

If we calculate the raw average rotational period, the obtained values from
Table \ref{table1} are 6.95, 6.88, and 6.75 hours for the entire sample, for
both the TNOs and the Centaurs only, respectively. On the other hand, if we
fit Maxwellian distributions to the rotational frequencies, the mean of the
distributions change slightly. The two main populations are separated here,
giving a mean rotational period of 7.35 hours (3.19 cycles/day) for the
entire sample (upper panel), 7.71 hours (3.11 cycles/day) for the TNOs
(center panel), and 8.95 hours (2.68 cycles/day) for the smaller Centaurs
(lower panel).

As we can see in Fig. \ref{fig04}, there appears to be some fast rotators.
In the center panel, for the TNOs, there is Ceto, 1996 TP$_{66}$, and 2004
GV$_9$ with single peak rotational periods of 2.21, 1.96, and 2.93 hours,
which are quite short and unrealistic. The reason for this is that all 3 of
the afore mentioned objects have light-curve amplitudes smaller than 0.15mag
and were considered to have single-peaked light curves, whereas in reality
they have double-peaked light curves. A similar case occurs for the Centaur
Chiron and 2002 GZ$_{32}$ (lower panel), which have a single peak rotational
period of 2.96 and 2.9 hours, respectively, for the same reason.

\section{Correlations}

To search for correlations between different physical and orbital parameters
(rotational period, peak-to-peak amplitude, semi-major axis, perihelion
distance, aphelion distance, eccentricity, inclination), we use
non-parametric statistical tests.

The presence of several periods and/or amplitudes for most of the objects,
encouraged us to develop a customized code that randomly chooses one
combination. The program created the individual combinations of period and
amplitude pairs and calculated the correlation between one of these
combinations (of period and amplitude) and the selected variable (e.g.,
colors, orbital parameters). The final choice of correlation, presented in
Table \ref{correl}, is the median among those correlations.

The statistical methods do not assume any particular population probability
distribution, nor any functional shape. In particular, we compute all the
correlations using the Spearman rank correlation, $\rho$
\citep{Spearman1904}. This method is distribution-free and less sensitive to
outliers than others methods. We studied the strength of the correlations by
computing the Spearman coefficient $\rho$ and the significance level (SL).
The $\rho$ coefficient has values between -1 and 1. A correlation may exists
if $\rho$ $>$ 0, while $\rho$ $<$ 0 implies a possible anti-correlation. A
zero value means that there is no correlation at all. The significance of
$\rho$ is measured by SL, the probability that the null hypothesis (samples
not correlated) is not true. We consider that correlations or
anti-correlations are strong when $\rho$ (in absolute value) is greater than
0.6, a possible weak correlation exists when $\rho$ values are between 0.3
and 0.6, and no correlations are present for values less than 0.3. The
following criteria were used: SL greater than 95\% = reasonably strong
evidence of correlation; SL greater than 97.5\% = strong evidence of
correlation; and SL greater than 99\% = very strong evidence of correlation.
The interpretation of $\rho$, SL, and the graphical representation of the
magnitudes allows us to detect possible correlations among the different
magnitudes. Apart from these criteria, we consider that a possible
correlation could exist when $\rho$ is greater than or equal to 0.3 and SL
is greater than 80\%, in particular for samples of only a few objects. We
emphasize that this particular threshold ($\rho$ $>$=0.3; SL$>$ 80\%)
provides only a hint of a possible correlation. We note that in the majority
of these cases, the sample is so small, that more observations are required
in some of the cases listed in Table \ref{correl}.

The Spearman statistical analysis of the results was made by separating the
dataset into different dynamical subgroups, i.e., classical objects, which
exhibit orbits with moderate eccentricities; resonant objects, trapped by
Neptune in mean motion resonances (the Plutinos being the most
representative population, in the 3:2 resonance), and scattered disk objects
(with orbits of high eccentricities and sometimes high inclinations, due to
close encounters with the planet Neptune in the past). The classical group
is sub-classified into two sub-groups: the so-called (dynamically) hot
classical objects with orbital inclinations $>$ 4.5$^o$, and the
(dynamically) cold classical objects, with inclinations $<$ 4.5$^o$.
Finally, the Centaurs represent TNOs scattered towards the inner Solar
System and they reside between the orbits of Jupiter and Neptune.

The analyzed parameters were the rotational period, light-curve amplitude,
colors (B-V, R-I and V-R), absolute magnitude H, and orbital parameters
(a,e,i,Q and q). It should be mentioned that a data set of colors was
presented in \cite{Santos-Sanz2009}. All parameters were tested against all
the other parameters in the entire sample and then in sub-populations. The
corresponding correlation coefficients and significance levels are listed in
Table \ref{correl}, where only the relevant cases are shown. We made the
test correlations with all the parameters in all the populations and then
listed only the correlations that were found to be significant.

For the case of amplitude versus H, values of $\rho$ and SL for all the
sample and in particular for the classical TNOs, clearly indicate that the
smaller (and collisionally evolved) objects are probably more elongated than
the larger ones. This could explain the larger amplitudes observed for
smaller objects. In the classical population (39 objects), the value of
$\rho$ = 0.47 and SL = 99.5 \% indicates a stronger correlation than the
entire population with $\rho$ = 0.40 and SL = 99.8 \%. For the Centaurs and
Plutinos, we do not find any evident correlation between these two
magnitudes.

The particular case of P versus B-V correlations may suggest that objects
with shorter rotation periods may have suffered more collisions (and thus
have bluer surfaces) than objects with longer ones. The collisions could
resurface the objects with fresh ices making them bluer and modifying the
spins toward faster rotation periods. There is a strong evidence of a
correlation in the classical subpopulation, although this sub-sample has
fewer than 20 objects. More color determinations of objects with known
periods would clearly be helpful, not only to confirm or exclude this
possible correlation but to confirm whether there is some color variation in
the surface of the body. The results obtained for the P versus H relations
of Centaurs (correlation) and Plutinos (anti-correlation) might be related
to the different mean sizes of the Centaur and Plutino populations and could
be related to a different primordial origin and/or evolution. In particular,
physical parameters such as the color B-V and rotational period are known
only for brighter and larger objects. We, again note the low number of
objects. We found some weak anti-correlations between the amplitudes and
orbital parameters Q and e, whose possible causes are not clear to us.

\begin{figure}
\resizebox{18cm}{!}{\includegraphics[angle=0]{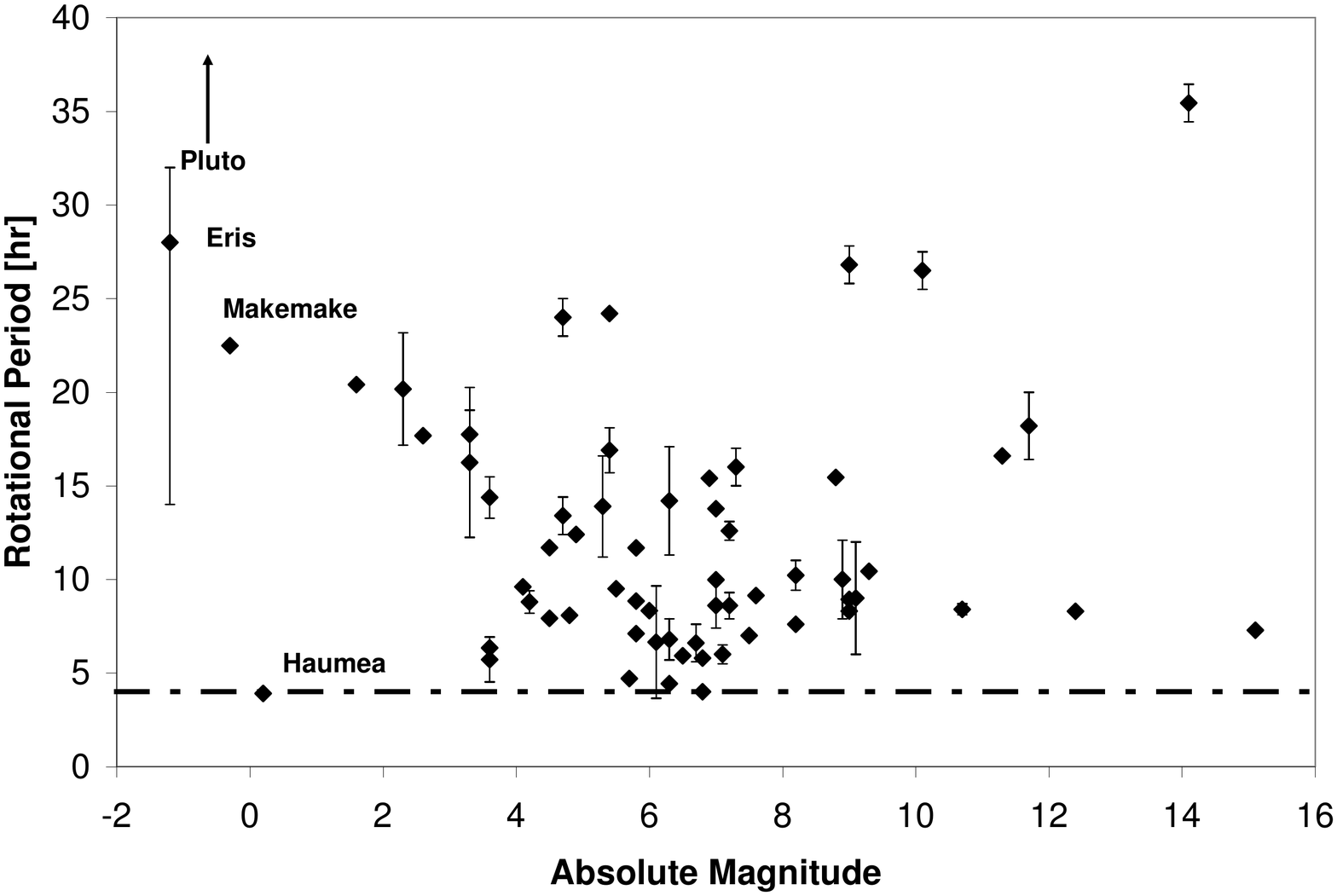}}
\caption{Rotational Period against absolute magnitude for all the objects.
The horizontal line at 4 hrs is the spin barrier mentioned in the text. }
         \label{fig05}
   \end{figure}

Finally, another interesting plot is the rotational period versus absolute
magnitude (an indicator of the size), where we can see (Fig. \ref{fig05})
that a rotation barrier could exist close to 4 hours. All the points have an
error bar that indicates the period determination error when the result is
unique and a larger error when the period has several values in the
literature. A typical example of multiple determinations in the literature
is Eris, for which the period has been determined to be either 13.7 h
\citep{Duffard2008} or 25.92 h \citep{Roe2008}.

\section{Densities and internal structures}

With the rotational period sample, it is possible to derive some constraints
on the density of the objects. Several authors have presented different
estimations on TNO densities. \cite{Pravec-Harris2000} derived a simple
expression that approximates the critical (minimum) period for a rotating
body as a function of density and light-curve amplitude. This relation
assumes a fluid body. \cite{Sheppard2008} used the periods and shapes
(derived from light curves) to identify an apparent trend of larger
(brighter) objects being denser. Finally, \cite{Tancredi2008} analyzed TNO
densities and internal structures to determine whether TNOs could be dwarf
planets.

We note that objects with radii r $>$ 200 km have presumably never been
disrupted by impacts \citep{DavisFarinella1997}, but have probably been in
some way fractured. If we assume that TNOs have relaxed to equilibrium
shapes, then their rotation states can be used to place limits on their
densities. The centripetal acceleration produced by self-gravity (bulk
density) must be sufficient to hold the material together against the
inertial acceleration due to rotation. Although TNOs consist of solid
material, their presumably fragmentary structure validates the fluid
approximation as a limiting case.

Densities have been proposed to be low, 600 kg/m$^3$ for 2001~QG$_{298}$ and
in some cases, covering a wider range to higher densities of 2600 kg/m$^3$
as in 2003 EL$_{61}$. The lower density is similar to that inferred for some
comets, which are objects two orders of magnitude smaller. Bulk densities
substantially below that of pure water ice are usually taken to imply an
ice-rich composition and porosity. It is expected that density variations
are not only related to intrinsic compositional differences, but might be
caused by variations in the porosity. On the other hand, for the dwarf
planet class, porosity cannot be important. It needs to be mentioned that a
density of 2600 kg/m$^3$ implies a rock/ice ratio of $\sim$85/15\% when the
internal structure is modelled \citep{McKinnon2008}.

\subsection{A simple Monte Carlo model}

Taking into account all the light curve amplitude and rotational period data
presented, it can be seen that ``flat curves" are dominant. As mentioned
before, this could be because we are observing large presumably round
objects. In the near future, more light curves should be obtained of smaller
objects and the question may then arise of whether this tendency remains and
how many non-flat curves are observed. To estimate how many non-flat curves
are expected in an observed population, we developed a simple model.

The first step in our model is to assume a Maxwellian rotational frequency
distribution such as that obtained in this work (Fig. \ref{fig04}), from the
observational data. Then, each object from a set of 100 000 of fixed density
is randomly assigned to a rotational period from the distribution. The
Maxwellian rotational frequency distribution adopted ranges from 8 to 0.5
cycles/day, with a bin of 0.1 and a peak at 3.11 cycles/day (7.71 hours), in
the observed distribution.

One important constraint of the model is that light curves are assumed to be
related to the shape of the body. The model aims to determine the percentage
of objects that create non-flat light curves due to elongated shapes. Other
causes of brightness variation (not assumed here) are, for example, eclipses
in binary objects or surface albedo variations.

For the model calculations, it is assumed that each object rotates in
hydrostatic equilibrium and the Chandrasekhar formalism
\citep{Chandrasekhar1987} is used. The questions that we try to answer is
how many objects (of a given density) with different rotational periods
could have an irregular shape with axes of lengths such that a$>$b$>$c (a
Jacobi ellipsoid), and how many objects acquire a MacLaurin shape (i.e.,
a=b$>$c). The final percentage of MacLaurin or Jacobi objects should be
compared with the observed data.

Jacobi ellipsoids produce a non-flat light curve if the ratio of the axis
lengths are greater than the photometric error. It is then necessary that
the rotating axes are not aligned along the observer line of sight.
MacLaurin objects will produce a flat light curve independently of the
rotational axis direction. In all cases the objects are considered at
opposition and in a simple rotation around the shortest axis. According to
\cite{Burns1973}, rotational excitation damping times are orders of
magnitude shorter than the age of the Solar System and are also orders of
magnitude shorter than the typical time between collisions in the
transneptunian region. In their study of ellipsoids, \cite{Molina2003}
emphasized that damping times can be even shorter if one takes into account
the energy dissipation caused by internal stresses. Thus, it is expected
that the transneptunian objects must be purely in rotation around their
principal axes of maximal moment of inertia, because the collisions that can
excite their rotations take far longer to occur than the typical rotational
excitation damping time. The TNOs have been found to be in pure rotation in
other works, e.g., \cite{Ortiz2003b, Lacerda-Luu2003}

For all objects in each run, the dimensionless parameters $\Omega$ and
$\Lambda$ (dimensionless angular velocity and dimensionless angular
momentum) \citep{Chandrasekhar1987} were calculated. From these pairs of
parameters and using Chandrasekhar's tables, it is possible to determine
whether the object has a Jacobi or MacLaurin figure or if a non-equilibrium
figure is possible. To obtain a complete sample of all possible densities,
the model was run for several density values between 500 and 3000 kg/m$^3$.

\subsection{Results for the model}

For each density, the percentage of Jacobi, MacLaurin, and non-equilibrium
figures were calculated and are presented in Table \ref{model}. In the case
of the Jacobi ellipsoids, there is an interval of allowed rotational periods
where the objects are in equilibrium. For example, for a density of 1000
kg/m$^3$, only 12.61\% of the objects are Jacobi ellipsoids and they rotate
with periods between 6.31 and 7.05 hours. For this same density, 55.63\% of
objects have a MacLaurin figure and the remaining 31.75\% are
non-equilibrium figures that were discarded from the sample.

\begin{figure}
\resizebox{18cm}{!}{\includegraphics[angle=0]{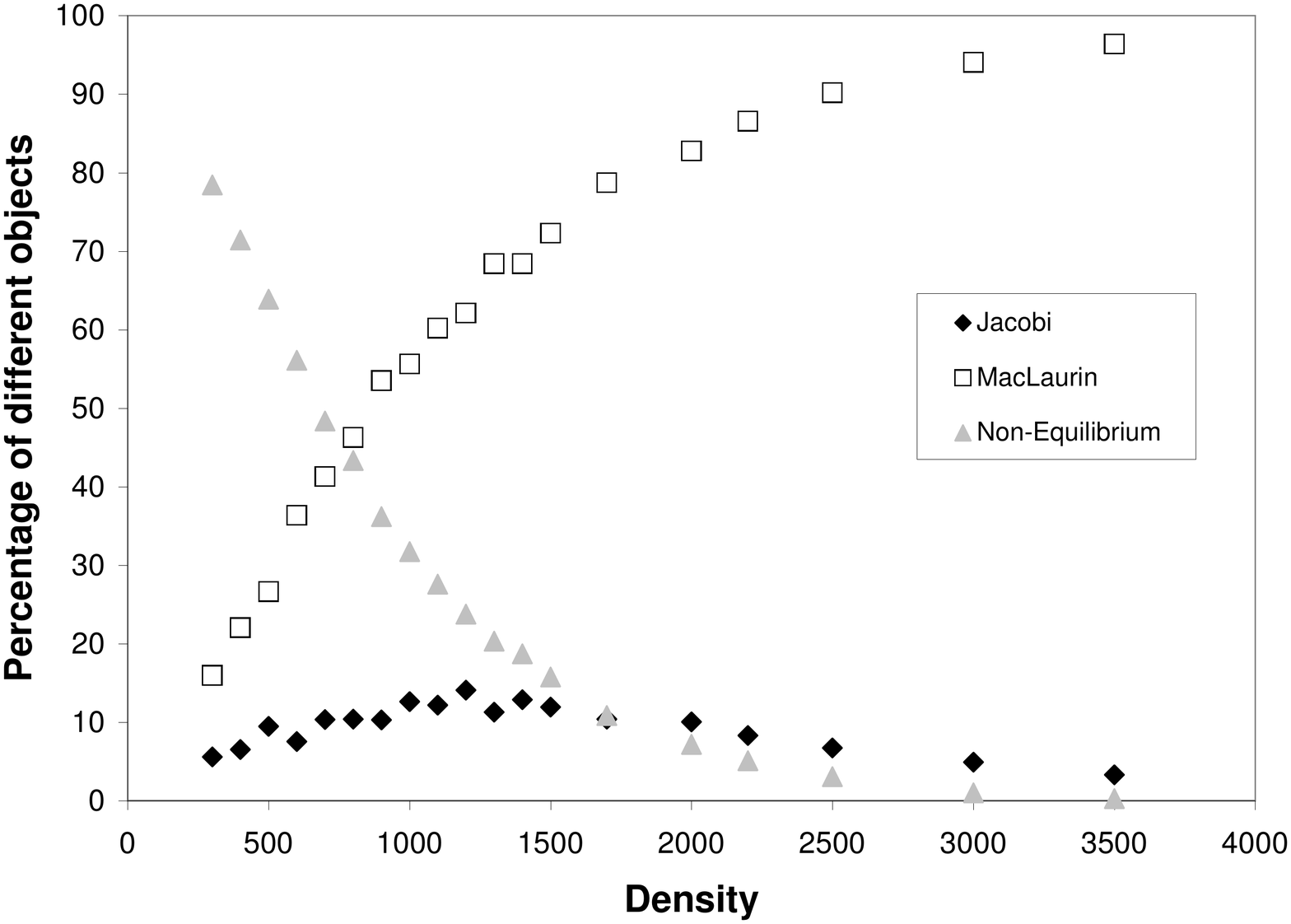}}
\caption{Percentage of the 3 different kinds of object produced by the model
against density (in kg/m$^3$). Jacobi ellipsoids are shown in black
diamonds, MacLaurin spheroids in white squares and finally triangles are the
non-equilibrium figures. }
         \label{fig06}
   \end{figure}

In Fig. \ref{fig06}, the percentage of Jacobi, MacLaurin, and
non-equilibrium objects is plotted against density for all the runs. As can
be seen, when the density increases the percentage of MacLaurin spheroids
and non-equilibrium figures, increase and decrease, respectively. The
percentage of Jacobi ellipsoids has a maximum close to the densities of
1200-1300 kg/m$^3$.

\begin{figure}
\resizebox{18cm}{!}{\includegraphics[angle=0]{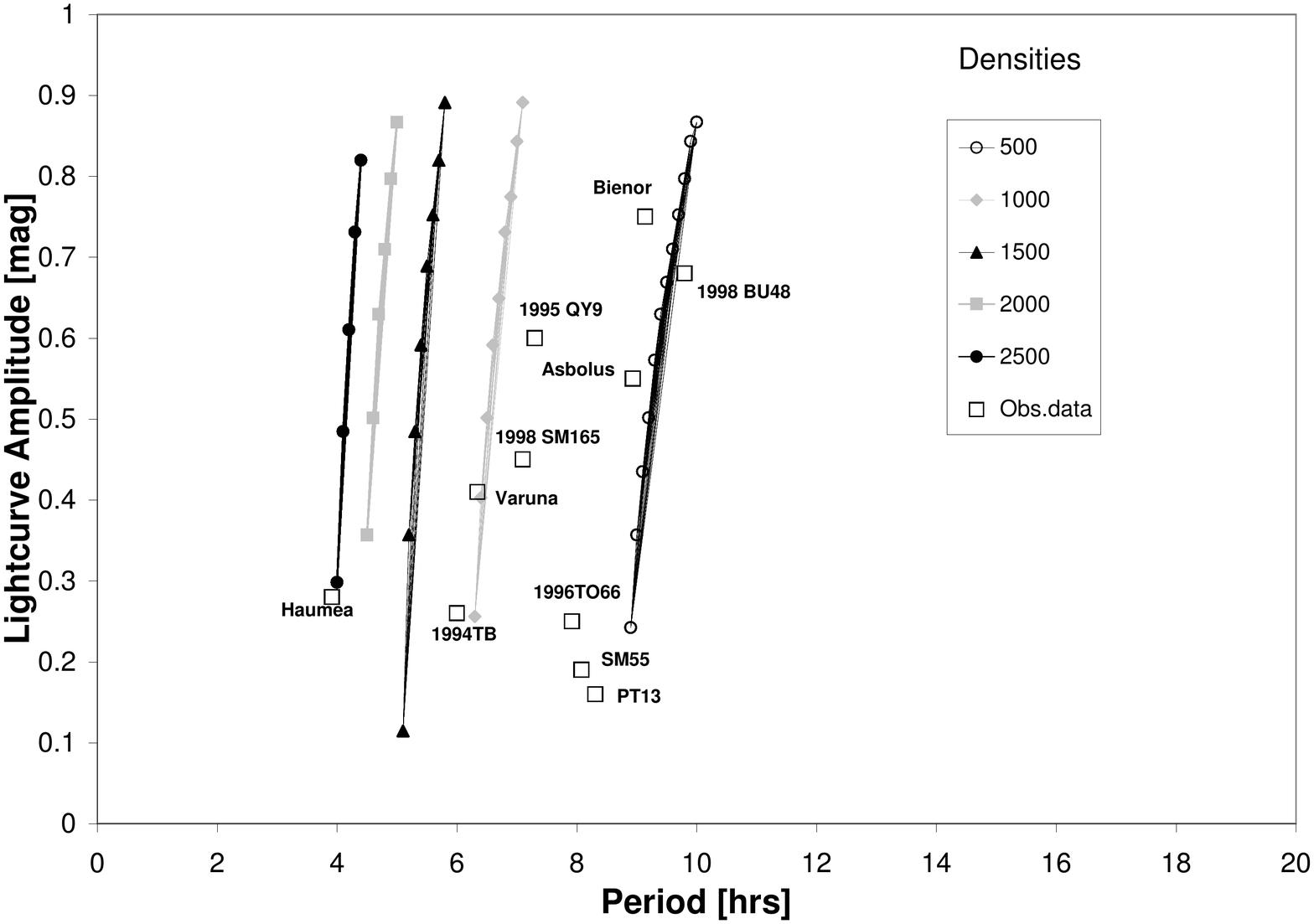}} \caption{light
curve amplitude vs. rotational period for several model runs compared with
some observational data. Densities are in kg/m$^3$}
         \label{fig07}
   \end{figure}

The amplitude of the light curve of each synthetic object is then calculated
from the axis ratio b/a of the Jacobi ellipsoid. For a fixed density, the
ratio b/a of the ellipsoid increases with the rotational period. The
obtained axis ratio was then modified by a random orientation of the
rotational axis. This means that an elongated Jacobi ellipsoid would be seen
pole-on and no light-curve variation would be observable. In Fig.
\ref{fig07}, the light curve amplitude for several runs of the model is
plotted against rotational period. As can be seen, for each density there is
an interval of allowed rotational periods for Jacobi ellipsoids. All the
period-lightcurve amplitude pairs below that curve are allowed after the
correction of the viewing angle. We note that the allowed rotational period
interval for a lower density is longer than that allowed for higher
densities, as expected. The rotational period interval becomes shorter when
the density increases, in other words, the curve becomes steeper as the
density increases.

Observed data is included in the figure to compare with the modelled data.
From this figure, the density of a real object can be estimated by assuming
that the objects are viewed equatorially. Haumea has a estimated density of
2500 kg/m$^3$, while most of the other selected objects have variations
between densities 500 to 1000 kg/m$^3$.

Finally, in Fig. \ref{fig08} the percentage of objects with different
figures is presented for several densities. MacLaurin figures were added to
the Jacobi ellipsoids in the case of small light curve amplitudes (0.1 mag).
Here it is assumed that if a light curve amplitude is detectable in a
MacLaurin object, it will be smaller than $\sim$ 0.1 mag, because it will be
produced by albedo features. For the different model runs (densities) we can
observe that the percentage of objects with small light curve amplitude
(0.1mag) is always higher than the percentage of the more elongated ones, as
expected. This percentage increases from 30\% for a low density such as 500
kg/m$^3$ to more than 90\% for a higher density like 2500 kg/m$^3$. On the
other hand, as can be seen, the percentage of objects with large amplitudes
is independent of density and is always smaller than 20\%. The percentage of
MacLaurin objects sharply increases with density and this is the main reason
for the differences between the curves at the left part of Fig. \ref{fig08}
(small light curve amplitudes).

In Fig. \ref{fig08}, the observational data are also plotted separated into
two intervals of absolute magnitudes. The percentage of objects in the
interval of H between [-1,5] and [-1,7] is taken from Table \ref{table1}.
The behavior of both curves are the same as for the modelled ones. A large
percentage of the objects have low amplitude light curves. Comparing this
behavior with the modelled curves it is possible to estimate the mean
density of the entire sample. As can be seen, the observational data agree
with the output of the corresponding model run for 1000 kg/m$^3$ density.
When smaller objects are included (black squares), the observed points are
reproduced most accurately with an even lower density.

At the other side of the plot, it can be seen that observational data are
lacking with respect to the model prediction. A blow-up of the plot for the
larger amplitudes is inserted in Fig. \ref{fig08}. As can be seen, some
objects with larger light curve amplitudes are expected to be found.

To summarize, if hydrostatic equilibrium is considered with a Maxwellian
rotational frequency distribution, a large percentage of ``flat light
curves" is obtained. This is because of the large percentage of MacLaurin
figures added to the Jacobi ellipsoid with small amplitude (due to a small
axis ratio or viewing angle). It is important to note that for the results
mentioned above, only a distribution of rotational frequencies and a fixed
density are assumed. No other physical variable is needed.

Finally, if a size distribution is included in the model, it is possible to
compare the objects with the real sizes. For that, a cumulative surface
density of TNOs was included as described in \cite{Petit2008} between 18 $<$
R $<$ 22 magnitude. Assuming a mean albedo of 0.12 and mean distance of 40
AU, the smallest objects in this group have a radius of 86 km. In other
words, hydrostatic equilibrium seems to be valid for objects smaller than
assumed before \citep{Sheppard2008}. This could be because the internal
cohesion forces are smaller in TNOs than in asteroids. An estimated value of
the tensile strength for this size that is overcome by the object's gravity,
allowing the objects to behave like a fluid, is obtained from equation 3 of
\cite{Thirouin2009}. The tensile strength is only 0.1 MPa. This small
tensile strength is much smaller than that of typical geophysical solids.

\section{Discussion}

With all the available data, we have compiled different histograms to see
the distribution of physical parameters such as rotational periods and
light-curve amplitudes. With this larger number of light curves, the results
obtained for the mean rotational period and peak-to-peak amplitudes are more
reliable than in previous studies. The criteria to separate flat light
curves from the others is important when selecting a single or double peak
rotational period. As is demonstrated, the distribution of periods infers a
lower limit (for a single-peak period) and upper limit (for double-peak
periods). The result of the combination of both periods is a more reliable
result. The threshold of 0.15mag for the transition from the albedo-marking
light curves to shape-induced light curves is somewhat arbitrary and all the
results are based on this value. If a light curve amplitude magnitude of
0.10mag is considered the obtained results are similar, reaching a mean of
7.84 hours and 7.12 hours for all the sample and for only TNOs,
respectively. The goodness of the Maxwellian fit is similar to the 0.15mag
case. However, if we use a threshold of 0.2mag, the fit is degraded.

Why are their surfaces so uniform? The sample of flat light curves is
dominated by larger objects. These bodies are expected to be round and in
hydrostatic equilibrium. In these bodies, it is natural to ask: could the
reason for flat light curves be the presence of a thin atmosphere or recent
outgassing/resurfacing?

The environment in the trans-neptunian region might favor the formation of
tenuous permanent or transient atmospheres. This idea is supported by
several pieces of evidence such as (i) the intrinsic and/or induced activity
on some bodies, such as in Chiron \citep{Duffard2002} and other Centaurs.
Activity has been reported for comets at large heliocentric distances and
much beyond the water ice sublimation limit \citep{Hainaut2000}. (ii)
Abundance of volatiles with sublimation temperatures comparable to the local
temperature. (iii) The most important, the atmospheres of Pluto
\citep{Yelle1997} and Triton \citep{Yelle1995}.

Our study is based on a photometric analysis, so we cannot determine the
presence of atmospheres. The observational result is that most TNOs have
small amplitudes. Taking the evidence together, it appears reasonable to
expect the formation of thin gaseous and/or dusty atmospheres in large TNOs,
because they would have the mass needed to avoid direct escape of materials
into space. There is a study by \cite{Lykawka2005} about the theoretical
presence of an atmosphere according to several criteria and the possibility
of the presence of a thin atmosphere in the studied bodies that takes into
account volatiles such as methane, nitrogen, and carbon monoxide, the most
common chemical species in the outer Solar System. It is noticeable that
only TNOs with a diameter of 700 km (albedo criteria are explained in that
article) can retain an atmosphere. The large self-gravity of the TNOs should
allow them to remain close to hydrostatic equilibrium, retain extremely
volatile ices, possibly have tenuous atmospheres, and be differentiated.
Thus, the surfaces as well as the interior physical characteristics of the
largest TNOs may differ significantly from those of the smaller objects.

A nearly uniform surface may be explained by an atmosphere that is frozen
onto the surface of the objects when close to aphelion. The atmosphere, such
as Pluto's, may become active when near perihelion, effectively resurfacing
itself every few hundred years. Taking into account all the known light
curves with known amplitudes and the work of \cite{Lykawka2005}, it is
possible to identify possible candidates to be spectroscopically observed by
searching for spectral signatures of an atmosphere.

The transient atmospheres and atmospheric freeze-out as a means of
resurfacing and homogenization would apply to the largest objects only. An
alternative scenario might be the collisional resurfacing, which may explain
other phenomena in the transneptunian belt
\citep{Gil-Hutton2002,Gil-Hutton2009}.

In the model presented here, for different runs of only one density, the
percentage of MacLaurin figures is always superior to 50\% for densities
higher than 900 kg/m$^3$. As can be seen in Fig. \ref{fig08}, the percentage
of observed objects with light curve amplitudes smaller than 0.2mag is
always 70\%. When compared to real data, it needs to be mentioned here that
this sample is biased as observers tend to publish light curves with large
amplitudes and easy-to-measure rotational periods.

According to our simple model, lower mean densities are obtained when we
analyze the sample with the largest size span, from H=[-1,7], compared to
the interval H=[-1,5]. Hence, it appears that the ``small" objects are also
in hydrostatic equilibrium, but have lower densities than the largest
objects. Therefore, an obvious refinement of our model could be to
incorporate a size-dependent density together with the size distribution of
the bodies whose brightness variability has been studied. Other improvements
in the model could be the use of non-random spin axis orientations to test
the effect of different spin axis distributions.

Despite the simplicity of our model, its results are compatible with
observations. Currently, the only input of the model is a Maxwellian
rotational frequency distribution, but the real distribution might differ
from a Maxwellian, since this function is representative of a collisional
equilibrium population. However, the true scenario might be represented by a
combination of a collisional evolved populations and a more primordial one.
Therefore, an additional improvement of the model would be the possibility
to include different rotational frequency distributions.

As a caveat to the densities derived here, we can mention the work of
\cite{Holsapple2001} and \cite{Holsapple2004} who determined the spin limits
of bodies using a model for solid bodies without tensile or cohesive
strength. That theory included the classical analysis of fluid bodies given
by MacLaurin, Jacobi, and others as a special case. For the general solid
bodies, it was shown that there exists a very wide range of permissible
shapes and spin limits, and explicit algebraic results for those limits were
given. \cite{Holsapple2007} extended these analyse to include
geological-like materials that also have tensile and cohesive strength. In
that paper, the author asked whether larger bodies such as the TNOs should
have relaxed fluid shapes or might have shapes that are allowable by
granular zero cohesion material? The important result is that if they are
not governed by fluid theory, the mass and shapes that have been inferred
could be substantially incorrect \citep{Holsapple2007}.

Finally, we must mention that the search for correlations between light
curve variables and orbital parameters is hampered by bias and selection
effects \citep{Binzel1989}, and our conclusions in this respect are weak.
However, the correlations between light curve parameters and physical
properties such as colors and size that we derived are less affected by bias
and we believe are reliable.

\section{Conclusions}

We have analyzed the light-curve data of \cite{Thirouin2009} and others in
the literature to derive light-curve parameters, which have been compiled
here. We have presented the most complete distribution of rotation periods
and amplitudes so far. For the rotation periods, the mean value obtained by
a Maxwellian fit in rotational frequencies is 7.71 hours (3.11 cycles/day)
considering only the TNOs, 8.95 hours (2.68 cycles/day) considering only
Centaurs, and 7.35 hours (3.19 cycles/day) considering all the sample
altogether. These results were obtained by taking into account the criteria
of considering a single peak light curve for those objects with amplitudes
lower than 0.15 mag and a double peak light curve for the more elongated
ones.

This mean rotational period is slightly longer than the mean value for the
largest Main Belt asteroids, 6.9 hours \citep{Binzel1989}, but we state that
this value may be affected by observational biases. It is important to
stress that in photometry devoted to the study of the rotational properties
of minor bodies, there exists and important observational bias towards
shorter periods and higher amplitude light curves. How this bias can be
estimated remains an open problem, but it certainly depends on the number of
objects for which a period is obtained. Some steps in this direction were
already taken by \cite{Thirouin2009}. Concerning other biases, if
observational time for larger telescopes is granted then smaller TNOs will
be accessible for the determination of the light curves, and the sample will
not only increase in size but the bias in size will become less important.

One of the most important correlations found in this work is the light curve
amplitude and absolute magnitude of all the sample of KBOs. This correlation
is not a surprise and is probably related to the most collisionally evolved
small population, which should be more elongated objects. On the other hand,
the less collisionally evolved population (larger and then brighter objects)
is the more numerous one in our sample ($\sim$60\%), so we caution the
reader that some of our results could be somewhat biased. Another
interesting correlation is the rotational period and B-V color, which could
indicate the age of the surface (younger surfaces being bluer). This
correlation appears to be consistent with the idea that the more
collisionally evolved objects spin faster and have bluer surfaces.

\begin{figure}
\resizebox{18cm}{!}{\includegraphics[angle=0]{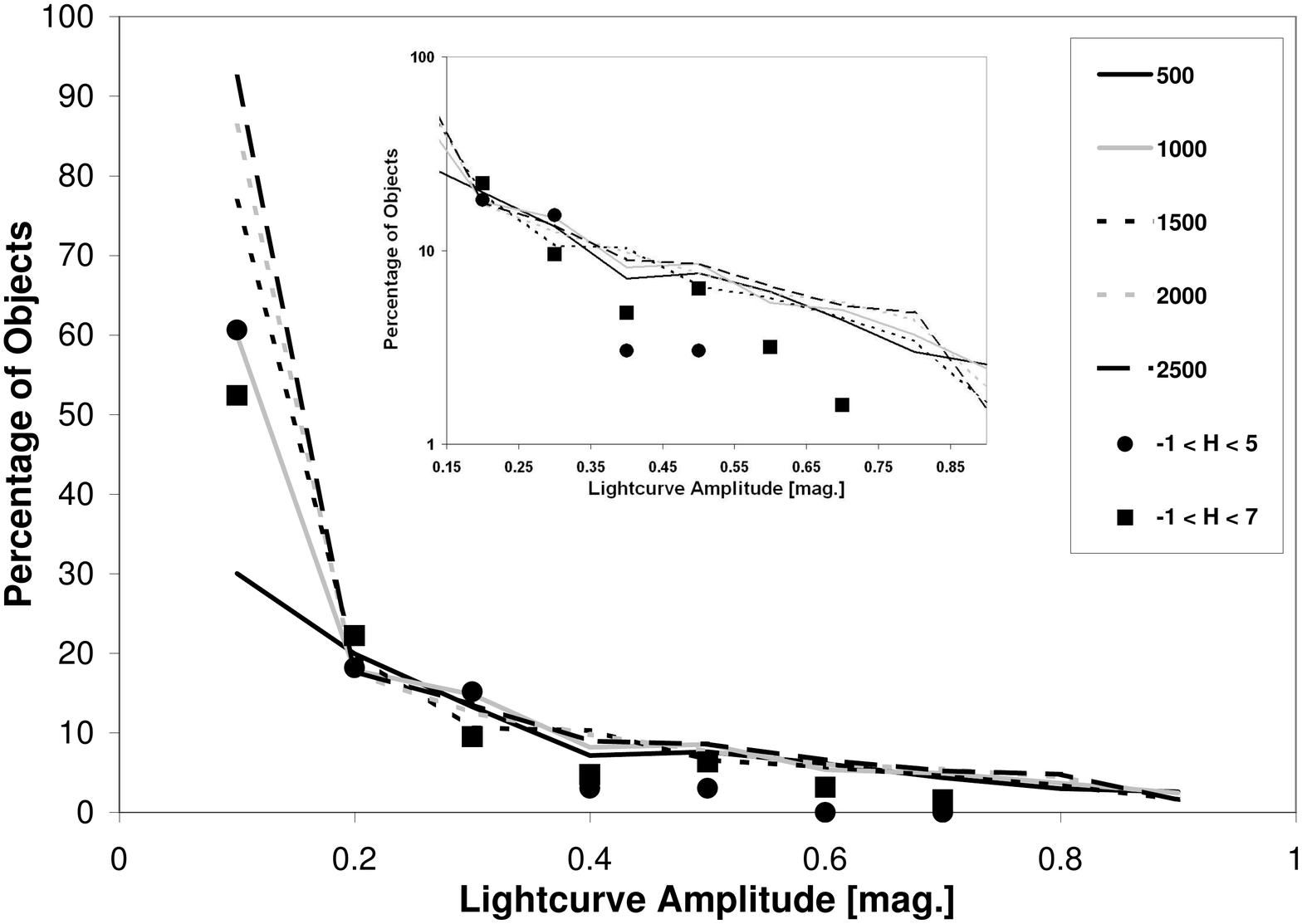}}
\caption{Percentage of the objects having a certain light curve amplitude
from the model for different densities. MacLaurin object and Jacobi
ellipsoid were added together only for a light curve amplitude of 0.1 mag.
Circles and squares are observational data using values of absolute
magnitude between [-1,5] and [-1,7], respectively. }
         \label{fig08}
   \end{figure}

A simple shape model has been developed that is able to reproduce some
results on the relative percentage of presence of MacLaurin/Jacobian
figures, when assuming hydrostatic equilibrium. In the entire sample, less
than 12\% of the objects in hydrostatic equilibrium have a Jacobi shape.
Most of the objects have MacLaurin figures that have small light curve
amplitudes (because of albedo marks). The model that best fits the
observations requires densities between 1000 to 1200 kg/m$^3$. We must
mention that as was highlighted in \cite{Thirouin2009} there is an
observational bias in the percentage of low amplitude light curves, which
means that perhaps 75\% of the objects have low light curve amplitudes. This
implies that the mean density is likely somewhat higher, around 1500
kg/m$^3$ (see Fig. \ref{fig08}).

For plausible albedos of 0.04 to 0.20, the absolute magnitude range H=[-1,7]
includes objects with diameters as small as $\sim$120 km. Those objects
would also be in hydrostatic equilibrium. We end with the provocative
thought that the great majority of observed TNOs would qualify to be dwarf
planets, because they appear to meet the hydrostatic equilibrium condition.

\begin{acknowledgements}

RD acknowledges financial support from the MEC (contract Juan de la Cierva).
This work was supported by contracts AYA2008-06202-C03-01,
AYA2005-07808-C03-01 and P07-FQM-02998. We thank the referee for his/her
constructive and helpful comments.

\end{acknowledgements}

\bibliographystyle{aa}
%\bibliography{12601bib}

% -----------------------------------------------------------------

\clearpage

\clearpage

\clearpage

\clearpage

\clearpage

\clearpage

\clearpage

\clearpage

\clearpage

% -----------------------------------------------------------------
\begin{scriptsize}
\longtab{1}{
\begin{longtable}{llcccccc}
\caption{\label{table1}List of TNOs whose variability has been studied.
When a multiple determination of the rotational period is present, the preferred period is marked in bold.}\\
\hline\hline Object & Designation  &  Single Peak  & Double Peak &
Amplitude & Absolute  & Ref $^\mathrm{b}$\\
  &         & Period [h]    & Period [h] &
[mag.]& magnitude$^\mathrm{a}$ & \\

\hline
\endfirsthead
\caption{continued.}\\

\hline\hline Object & Designation  &  Single Peak  & Double Peak &
Amplitude & Absolute  & Ref $^\mathrm{b}$\\
  &         & Period [h]    & Period [h] &
[mag.]& magnitude$^\mathrm{a}$ & \\ \hline
\endhead
\hline
\endfoot

(134340) Pluto    &                   & 153.6           & -               & 0.33              & -0.7  & B97,TH90 \\
Charon            &                   & 153.6           & -               & 0.08              & 0.9   & B97 \\
(148780) Altjira  & 2001~UQ$_{18}$         &        -        &    -            &   $<$0.10         & 5.6   & S07 \\
(55576)~Amycus    & 2002~GB$_{10}$         & 9.76            &  19.52          & 0.16$\pm$0.01     & 7.8   & Th09 \\
(8405) Asbolus    & 1995~GO           & 4.47            & 8.9351$\pm$0.003& 0.55              & 9.0   & D98a,K00 \\
                  &                   & -               & -               & 0.34              & ...   & RT99 \\
(54598) Bienor    & 2000~QC$_{243}$        & 4.57$\pm$0.02   & 9.14$\pm$0.04   & 0.75$\pm$0.09     & 7.6   & O03b \\
                  &                   & 9.17            & -               & 0.34$\pm$0.08     & ...   & R07a \\
(66652) Borasisi  & 1999~RZ$_{253}$        & -               &-                & $<$0.05           & 5.9   & LL06 \\
(65489) Ceto      & 2003~FX$_{128}$        &  -              & 4.43$\pm$0.03   & 0.13$\pm$0.02     & 6.3   & SJ02 \\
(19521) Chaos     & 1998~WH$_{24}$         & -               &-                & $<$0.10           & 4.9   & SJ02,LL06 \\
(10199) Chariklo  & 1997~CU$_{26}$         & -               & -               & $<$0.1            & 6.4   & LL06 \\
(2060) Chiron     & 1977~UB           & -               & 5.9180 $\pm$ 0.0001  & 0.088$\pm$0.003& 6.5 & B89 \\
(83982) Crantor   & 2002~GO$_{9}$          & (6.97 or 9.67)$\pm$0.04& -        & 0.14$\pm$0.04     &  9.1  & O03b \\
                  &                   & -               & -               & 0.34              & ...   & RT99 \\
(60558) Echeclus  & 2000~EC$_{98}$         & 13.401          & 26.802          & 0.24$\pm$0.06     & 9.0   & R03 \\
(31824) Elatus    & 1999~UG$_{5}$          & 13.25           & -               & 0.24              &  10.1 & G01 \\
                  &                   & 13.41$\pm$0.04  & -               & 0.102$\pm$0.005   & ...   & B02 \\
(136199) Eris     & 2003~UB$_{313}$        & 13.69/28.08/32.13& -              & $<$0.1$\pm$0.01   & -1.2  & D07 \\
                  &                   & 3.55?           & -               & $\sim$0.05        & ...   & L07 \\
                  &                   & -               & -               & $<$0.01           & ...   & R07a,S07 \\
                  &                   & $~$25.92        & -               & $~$0.1            & ...   & R08 \\
(136108) Haumea   & 2003~EL$_{61}$         & -               & 3.9154$\pm$0.0002 & 0.28$\pm$0.04   & 0.2   & R06\\
                  &                   & -               & 3.9155$\pm$0.0001 & 0.29$\pm$0.02   &  ...  & LJ08 \\
                  &                   & 1.96            & 3.92              & 0.23 $\pm$0.02  &  ...  & Th09\\
(10370) Hylonome  & 1995~DW$_{2}$          &   -             & -               & $<$0.04           & 8.0   & RT99 \\
(38628) Huya      & 2000~EB$_{173}$        & 6.68/\textbf{6.75}/6.82 & -       & $<$0.1            & 4.7   & O03b \\
                  &                   &  -              & -               & $<$0.06           &... & O03b,SJ02,S02 \\
                  &                   &  -              & -               & $<$0.04           &...    & SJ03,LL06 \\
(28978) Ixion     &  2001~KX$_{76}$        & -               & -               & $<$0.05           &  3.2  & O03b,SJ03 \\
                  &                   & -               & -               & $<$0.10           & ...   & SJ02,LL06 \\
(136472) Makemake & 2005~FY$_{9}$          &11.24$\pm$0.01   & 20.54/\textbf{22.48} & 0.03$\pm$0.01& -0.3  & O07 \\
                  &                   & 7.65            & 15.30           & 0.014$\pm$0.002   & ...   & Th09 \\
(7066) Nessus     & 1993~HA$_{2}$          & -               &   -             & 0.5               & 9.6   & RT99 \\
                  &                   & -               & -               & $<$0.2            & ...   & D98b \\
(52872)~Okyrhoe   & 1998~SG$_{35}$         & 4.86/6.08       & 9.72/12.16      &  0.07$\pm$0.01    & 11.3  & Th09 \\
(90482) Orcus     & 2004~DW  & 7.09/\textbf{10.47$\pm$0.01}/17.43 &  \textbf{20.94} & 0.03$\pm$0.01 & 2.3 & O06 \\
                  &                   & 13.19           & -               & 0.18              &...    & RO7a \\
                  &                   &    -            & -               &  $<$0.03          &...    & S07 \\
                  &                   & 10.47           &  20.94          & 0.03$\pm$0.01     & ...   & Th09 \\
(5145) Pholus     & 1992~AD           & -               & 9.98            & 0.15/0.60   & 7.0   &B92,H92,F01,T05 \\
                  &                   & -               &   -             & 0.15              & ...   & RT99 \\
(50000) Quaoar    & 2002~LM$_{60}$         & -               & 17.6788$\pm$0.0004 & 0.13$\pm$0.03  & 2.6   & O03b \\
                  &                   & 8.84            & -               & 0.18$\pm$0.10     & ...   & R07a \\
                  &                   & 9.42            & 18.84           &  $\sim$0.3        & ...   & L07 \\
                  &                   & 8.84            & 17.68           & 0.13$\pm$0.04     & ...   & Th09  \\
(90377) Sedna     & 2003~VB$_{12}$         & 10.273$\pm$0.003& -               & 0.02              & 1.6   & G05 \\
(88611) Teharonhiawako &2001~QT$_{297}$    & -               & -               & $<$0.15           & 5.5   & Os03 \\
(32532) Thereus   & 2001~PT$_{13}$         & 4.1546$\pm$0.0001& 8.3091$\pm$0.0001 & 0.16$\pm$0.02  & 9.0   & O03b \\
                  &                   & -               & 8.34            & 0.16              &...    & FD03 \\
                  &                   & -               & 8.34            & 0.34$\pm$0.08     &...    & R07a \\
                  &                   & -               & 8.4             & 0.15              &...    & F01 \\
(42355) Typhon    & 2002~CR$_{46}$         & 9.67             & 19.34          & 0.06$\pm$0.01      & 7.2   & O03b \\
                  &                   & -               & -               & $<$0.05           & ...   & SJ03 \\
                  &                   & $>$5            &  -              &    -              & ...   & D08 \\
                  &                   & 9.67            & 19.34           & 0.06$\pm$0.01     & ...   & Th09 \\
(20000) Varuna    & 2000~WR$_{106}$        & 3.1718          & 6.3436$\pm$0.0001 & 0.43$\pm$0.01   & 3.6   & O03b \\
                  &                   &  -              & 6.34            &  0.42$\pm$0.03    & ...   & SJ02 \\
                  &                   & 3.17            & 6.34            &     0.5           & ...   & F01 \\
                  &                   &     -           &6.34358$\pm$0.00002 & 0.42           & ...   & B06 \\
                  &                   &      -          & 6.344           & 0.49$\pm$0.17     & ...   & R07a \\
                  &                   & 3.1709          & 6.3418          &  0.43$\pm$0.01    & ...   & Th09 \\
(15789)           & 1993~SC           &    7.7          &   -             & 0.04              & 6.9   & T97 \\
                  &                   &  -              & -               & $<$0.04           & ...   & RT99 \\
(15820)           & 1994~TB           & 3.0/3.5         &6.0/7.0          & 0.26/0.34         & 7.1   & RT99 \\
                  &                   & -               & -               & $<$0.04           & ...   & SJ02 \\
(19255)           & 1994~VK$_{8}$          & 3.9/4.3/4.7/5.2 & 7.8/8.6/9.4/10.4& 0.42              & 7.0   & RT99 \\
                  &                   & 4.75            & -               &    -              & ...   & CB99 \\
(24835)           & 1995~SM$_{55}$         & 4.04$\pm$0.03   & 8.08$\pm$0.03   & 0.19$\pm$0.05     & 4.8   & SJ02 \\
(32929)           & 1995~QY$_{9}$          & $~$3.5          & $~$7.0          & 0.60              & 7.5   & RT99 \\
                  &                   & Between 3.3 and 3.7 & -           & 0.60              & ...   & RT99 \\
                  &                   &    -            & 7.3$\pm$0.1     & 0.60$\pm$0.04     & ...   & RT99,SJ02 \\
(26181)           & 1996~GQ$_{21}$         &  -              &  -              &  $<$0.10          & 5.2   & SJ02\\
(15874)           & 1996~TL$_{66}$         & 12.1            & -               & $<$0.12           & 5.4   & O06 \\
                  &                   & -               & -               & $<$0.06           & ...   & LJ98,RT99 \\
                  &                   &$~$6/8.04/\textbf{$~$12}& \textbf{$~$24}  & 0.09$\pm$0.02 & ... & Th09 \\
(19308)           & 1996~TO$_{66}$  & \textbf{3.96$\pm$0.04}& \textbf{7.92$\pm$0.04}/5.9/9.6 & 0.25$\pm$0.05& 4.5& SJ03\\
                  &                   & -               & 11.9            & 0.25$\pm$0.05     & ...   & B06 \\
                  &                   &   -             & 6.25$\pm$0.03   & 0.12 to 0.33      & ...   & H00 \\
                  &                   &   -             &  -              &  $<$0.10          & ...   & RT99 \\
(15875)           &  1996~TP$_{66}$        &  1.96           & -               &  $<$0.04          & 6.8   & CB99 \\
                  &                   &  -              & -               & $<$0.12           & ...   & RT99 \\
(118228)          &  1996~TQ$_{66}$        & -               & -               & $<$0.22           & 7.1   & RT99 \\
                  & 1996~TS$_{66}$         & -               & -               & $<$0.14           & 6.5   & LL06 \\
                  &                   &  -              & -               & $<$0.16           & ...   & RT99 \\
(79360)           & 1997~CS$_{29}$         & -               & -               &  $<$0.08          & 5.1   & SJ02 \\
                  &                   & -               & -               & $<$0.22           &  ...  & RT99 \\
                  & 1997~CV$_{29}$         &  -              &  15.8       & 0.4         & 7.3   & CK04 \\
(33128)           & 1998~BU$_{48}$         & (4.9 or 6.3)$\pm$0.1& (9.8 or 12.6)$\pm$0.1 & 0.68$\pm$0.04 & 7.2  & SJ02 \\
(91133)           & 1998~HK$_{151}$        &  -              & -               & $<$0.15           & 7.6   & SJ02 \\
(26308)           & 1998~SM$_{165}$        & -               & 7.1$\pm$0.1     & 0.45$\pm$0.03     & 5.8   & SJ02 \\
                  &                   & 3.983           & -               & 0.56              & ...   & R01 \\
(35671)           & 1998~SN$_{165}$        & -               & 8.84            & 0.16$\pm$0.01     & 5.8   & LL06 \\
                  &                   & 5.03            & -               & 0.15              &...    & P02 \\
(33340)           & 1998~VG$_{44}$         & -               & -               & $<$0.10           & 6.5   & SJ02 \\
                  & 1998~XY$_{95}$         & -               & -               &  $<$0.1           & 6.2   & CB01 \\
                  &                   & 1.31            & -               &  $\sim$0.1        & ...   & CB01 \\
(26375)           & 1999~DE$_{9}$          & $>$12?          & -               & $<$0.10           & 4.7   & SJ02 \\
(79983)           &  1999~DF$_{9}$         &  \textbf{3.3} &  \textbf{6.65}/9.2 & 0.40$\pm$0.02    & 6.1   & LL06 \\
(40314)           & 1999~KR$_{16}$ & (5.840 or 5.929)$\pm$0.001& (11.680 or 11.858)$\pm$0.002 & 0.18$\pm$0.04& 5.8& SJ02 \\
(47171)           & 1999~TC$_{36}$         & 6.21            & -               & 0.06              & 4.9   & O03b \\
                  &                   & -               & -               & $<$0.07           & ...   & LL06 \\
                  &                   & -               & -               & $<$0.05           & ...   & SJ03 \\
(29981)           & 1999~TD$_{10}$         & 7.71$\pm$0.02   & -               & 0.65$\pm$0.05     & 8.8   & O03b \\
                  &                   & -               & 15.45           & 0.65              & ...   & R03,C03 \\
                  &                   & -               & 15.382          & 0.41$\pm$0.08     & ...   & M04 \\
                  &                   & -               & 15.3833         & 0.53$\pm$0.03     & ...   & RO5a \\
                  &                   & 5.8             &   -             &  0.65$\pm$0.05    & ...   & C00 \\
                  & 1999~TZ$_{1}$          &  -              & 10.438          & $<$0.1            & 9.3   & M08 \\
                  &                   &  ~5             &        -        & 0.15$\pm$0.02     & ...   & D08 \\
                  &                   & 5.211           &  10.422         &  0.06$\pm$0.01    &...    & Th09 \\
(80806)           & 2000~CM$_{105}$        & -               &-                & $<$0.14           & 6.3   & LL06 \\
                  & 2000~CP$_{104}$        &  -              &        -        & 0.06              & 6.7   & R08 \\
                  & 2000~FV$_{53}$         & 3.79            &  -              & 0.07$\pm$0.02     & 8.2   & TB05 \\
                  &                   & -               &  7.5            & 0.07              & ...   & TB05 \\
                  & 2000~CP$_{104}$        &  -              &    -            &  0.06             & 6.7   & R08 \\
(47932)           & 2000~GN$_{171}$        & -               & 8.329$\pm$0.05  & 0.61$\pm$0.03     & 6.0   & SJ02 \\
                  &                   &  -              & 8.329$\pm$0.05  & 0.60$\pm$0.03     & ...   & D08 \\
(82075)           & 2000~YW$_{134}$        &-                & -               & $<$0.10           & 5.0   & SJ03 \\
(150642)          & 2001~CZ$_{31}$         &     -           & \textbf{4.71}/5.23 & 0.21$\pm$0.02  & 5.7   & LL06  \\
                  &                   &  -              &  -              & $<$0.20           & ...   & SJ02 \\
(82158)           & 2001~FP$_{185}$        &-                &  -              & $<$0.06           & 6.1   & SJ03 \\
(82155)           & 2001~FZ$_{173}$        & -               &   -             & $<$0.06           & 6.2   & SJ02 \\
                  & 2001~KD$_{77}$         & -               & -               & $<$0.07           & 5.8   & SJ03 \\
                  & 2001~QF$_{298}$        &-                &-                & $<$0.12           & 4.7   & SJ03 \\
(139775)          & 2001~QG$_{298}$        & 6.89$\pm$0.0002& 13.7744$\pm$0.0004 & 1.14$\pm$0.04   & 7.0   & SJ04 \\
(88611B)          & 2001~QT$_{297B}$       & 4.75            & -               & 0.6               &  5.5  & Os03 \\
                  &                   & 5.50$\pm$0.02   & -               &   -               &  ...  & K06 \\
(42301)           &  2001~UR$_{163}$       & -               & -               & $<$0.08           & 4.2   & SJ03 \\
(126154)          & 2001~YH$_{140}$        & 6.22/\textbf{8.45$\pm$0.05}/12.99  & 12.38/\textbf{16.80}/26.72  & 0.16$\pm$0.05& 5.4   & O06 \\
                  &                   &      -      &     13.25            &  0.21$\pm$0.04    & ...   & S07 \\
                  &                   &  6.19/8.40/\textbf{13.36} & 12.38/16.80/\textbf{26.72} & 0.16$\pm$0.05 & ... & Th09 \\
(55565)           & 2002~AW$_{197}$ & 6.49/\textbf{8.87$\pm$0.01}/13.94/15.82  & 17.74  & 0.04$\pm$0.01 & 3.3  & O06 \\
                  &                   & 8.78             &   17.56        &  0.04$\pm$0.01     & ... & Th09  \\
(95626)           & 2002~GP$_{32}$         & $\sim$3.3       & $\sim$6.6       & $>1$              & 6.7   & K06 \\
                  &                   &    -            &  -              & $<$0.03           & ...   & S07 \\
(95626)           & 2002~GZ$_{32}$         &   -             & 5.80$\pm$0.03   & 0.15$\pm$0.03     & 6.8   & D08 \\
(73480)           & 2002~PN$_{34}$ & (4.23 or 5.11)$\pm$0.03 &(8.45 or 10.22)$\pm$0.06 & 0.18$\pm$0.04    & 8.2  & O03b \\
(55636)           & 2002~TX$_{300}$ &(\textbf{8.12} or 12.10)$\pm$0.08& (\textbf{16.24} or 24.20)$\pm$0.08& 0.08$\pm$0.02 & 3.3  & SJ03 \\
                  &                   &  7.89$\pm$0.03  & 15.78           & 0.09$\pm$0.08     & ...   & O04 \\
                  &                   & 4.08            & 8.16             &  0.04$\pm$0.01   & ...   & Th09 \\
(55637)           &     2002~UX$_{25}$     &7.19 or 8.4      & 14.38           & 0.21$\pm$0.06     & 3.6   & R05b \\
                  &                   & -               & 16.782          &  0.13$\pm$0.09    & ...   & RO5a \\
                  &                   &-                &-                & $<$0.06           &...    & SJ03  \\
(555638)          & 2002~VE$_{95}$         & 6.76/6.88/7.36/9.47  & -          & 0.08$\pm$0.04     & 5.3   & O06 \\
                  &                   & -               & -               & $<$0.06           & ...   & SJ03 \\
                  &                   & 4.90/6.18/\textbf{9.97}/17.23 &   9.80/12.36/\textbf{19.94}/34.46& 0.04$\pm$0.01 & ... & Th09\\
(136204)          & 2003~WL$_{7}$          &  8.24           &  16.48      & 0.04$\pm$0.01         & 8.7  & Th09 \\
                  & 2005~CB$_{79}$         & 6.76            & 13.52       & 0.13$\pm$0.02         & 5.0  & Th09 \\
(145451)          & 2005~RM$_{43}$         &  6.71           & 13.42       & 0.04$\pm$0.01         & 4.4   & Th09  \\
(145452)           & 2005~RN$_{43}$        & 5.62 or 7.32    & 11.24 or 14.64 & 0.04$\pm$0.01      & 3.9 & Th09 \\
(145453)           & 2005~RR$_{43}$        &  7.87           &  15.74   & 0.06$\pm$0.01            & 4.0& Th09 \\
(145486)           & 2005~UJ$_{438}$       & 4.16            & 8.32     & 0.11$\pm$0.01            & 10.5  &  Th09 \\
                   & 2007~UL$_{126}$ or 2002~KY$_{14}$ & 3.56 or 4.20   & 7.12 or 8.40 & 0.10$\pm$0.01    & 9.4 &  Th09\\
(120347)          & 2004~SB$_{60}$         & 6.09 or 8.01    & 12.18 or 16.02 & 0.03$\pm$0.01       & 4.4 & Th09 \\
(144897)          & 2004~UX$_{10}$         & 5.68            & 11.36         & 0.08$\pm$0.01       & 4.7 & Th09 \\
(119979)          &  2002~WC$_{19}$        & -               &-                &  $<$0.03          & 5.1   &  S07 \\
                  & 2003~AZ$_{84}$         & 4.32/5.28/6.72/\textbf{6.76$\pm$0.01} & - & 0.14$\pm$0.03 & 3.6  &O06 \\
                  &                   &    -            & 5.72$\pm$0.05   & 0.14$\pm$0.03     & ...   & SJ02  \\
                  &                   &  6.79           & 13.58           & 0.10$\pm$0.01     & ...   &Th09 \\
(120061)          & 2003~CO$_{1}$          & 4.51            & 9.02            & 0.06$\pm$0.01     & 8.9   & Th09 \\
                  & 2003~BF$_{91}$         & \textbf{9.1}/7.3 & -              &  1.09$\pm$0.25    & 11.7  & TB05 \\
                  & 2003~BG$_{91}$         & \textbf{4.2}/4.5/4.6/4.9 & -      & 0.18$\pm$0.75     & 10.7  & TB05 \\
                  & 2003~BH$_{91}$         &  -              &  -              & 0.42              & 11.9  & TB05 \\
                  &                   &  2.8            &   -             & $<$0.15           & ...   & TB05 \\
(120132)          & 2003~FY$_{128}$        &         -       & -               &  $<$0.08          & 5.0   & S07 \\
                  &                   &   $>$7          &  -              &    -              & ...   & D08 \\
                  &                   &  8.54           &  17.08          & 0.12$\pm$0.01     & 5.0 & Th09  \\
(120178)          & 2003~OP$_{32}$         & 4.845$\pm$0.03  &  -            & 0.26$\pm$0.04       & 4.1   & Ra08 \\
                  &                   &  4.05           &  8.10         & 0.10$\pm$0.01       & ...   & Th09 \\
                  & 2003~QB$_{112}$        &     -           &  -              & 0.46              & 7.0   & R08 \\
                  & 2003~QY$_{90A}$        & 3.4$\pm$1.1     & -               & 0.34$\pm$0.06     & 6.3   & KO6a \\
                  & 2003~QY$_{90B}$        & 7.1$\pm$2.9     &-                &  0.90$\pm$0.18    & 6.3   & K06a \\
                  & 2003~QY$_{111}$        &   -             &  -              & 0.60              & 6.6   & R08 \\
(174567)          & 2003~MW$_{12}$         & 5.90/7.87       &     11.80/15.74 & 0.06$\pm$0.01     &  3.6  & Th09 \\
(84922)           & 2003~VS$_{2}$          & 3.71 or 4.39    &  7.42           & 0.17$\pm$0.01     & 4.2   & O06,Th09 \\
                  &                   &    -            & 7.41$\pm$0.01   & 0.21$\pm$0.02     & ...   & S07 \\
(90568)           & 2004~GV$_{9}$          &     -           & -               & $<$0.1            &  4.0  & S07 \\
                  &                   &  -              & 5.86$\pm$0.03   & 0.16$\pm$0.03     &  ...  & D08 \\
(120348)          & 2004~TY$_{364}$        & 5.85$\pm$0.01   & 11.7$\pm$0.01   & 0.22$\pm$0.02     & 4.5   & S07 \\
\end{longtable}}
\end{scriptsize}
\begin{list}{}{}
\item[$^{\mathrm{a}}$]MPC values. \item[$^{\mathrm{b}}$] References:
B97:\cite{Buie1997}; D98a:\cite{Davies1998a};
K00:\cite{Kern2000};RT99:\cite{Romanishin-Tegler1999};
O03b:\cite{Ortiz2003b}; R07a:\cite{Rabinowitz2007a};
LL06:\cite{Lacerda-Luu2006}; D08:\cite{Dotto2008};
SJ02:\cite{Sheppard-Jewitt2002}; B89:\cite{Bus1989};
R03:\cite{Rousselot2003}; G01:\cite{Gutierrez2001}; B02:\cite{Bauer2002};
D07:\cite{Duffard2008}; L07:\cite{Lin2007}; S07:\cite{Sheppard2007};
R08:\cite{Roe2008}; S02:\cite{Schaefer2002};
SJ03:\cite{Sheppard-Jewitt2003}; B03:\cite{Bauer2003}; O06:\cite{Ortiz2006};
B92:\cite{Buie1992}; F01:\cite{Farnham2001}; T05:\cite{Tegler2005};
H92:\cite{Hoffmann1992}; O03a:\cite{Ortiz2003a}; G05:\cite{Gaudi2005};
Os03:\cite{Osip2003}; FD03:\cite{Farnham-Davies2003};
B06:\cite{Belskaya2006}; T97:\cite{Tegler1997};
CB99:\cite{Collander-Brown1999}; LJ98:\cite{Luu-Jewitt1998};
H00:\cite{Hainaut2000}; CK04:\cite{Chorney-Kavelaars2004};
R01:\cite{Romanishin2001}; P02:\cite{Peixinho2002};
CB01:\cite{Collander-Brown2001}; C03:\cite{Choi2003};
M04:\cite{Mueller2004}; R05a:\cite{Rousselot2005a};
C00:\cite{Consolmagno2000}; M08:\cite{Moullet2008};
TB05:\cite{Trilling-Bernstein2006}; R06:\cite{Rabinowitz2006};
SJ04:\cite{Sheppard-Jewitt2004}; K06b:\cite{Kern2006b};
O04:\cite{Ortiz2004}; R05b:\cite{Rousselot2005b};
LJ08:\cite{Lacerda-Jewitt2008}; Ra08:\cite{Rabinowitz2008};
K06a:\cite{Kern2006a}; O07:\cite{Ortiz2007}; D98b:\cite{Davies1998b};
D01:\cite{Davies2001}; TH90:\cite{Tholen1990}; Th09: \cite{Thirouin2009}.
\end{list}

\begin{table}
\caption{Most important correlations found using the light curve parameters
and orbital/physical variables as described in the text. The parameters
presented are the light curve amplitude (Ampl.), the perihelion (q) and
aphelion (Q) distances, inclination (i), eccentricity (e) and absolute
magnitude (H).} \label{correl} \centering
\begin{tabular}{llccc}
\hline\hline \noalign{\smallskip}
Population     & Correlated Magnitudes  & $\rho$ & SL(\%) & n \\
\hline \noalign{\smallskip}
All KBOs & Ampl. vs. Q & -0.22  & 96.96 & 100 \\
          & Ampl. vs. H &  0.35  & 99.94 & 100 \\
\hline
Classical  & Period vs. B-V   & 0.66  & 98.49 & 17 \\
           & Ampl. vs. Q      & -0.36 & 97.95 & 42 \\
           & Ampl. vs. H      & 0.49  & 99.80 & 42 \\
           & Ampl. vs. i      & -0.37 & 98.16 & 42 \\
           & Ampl. vs. e      & -0.34 & 96.74 & 42 \\
\hline
Hot        & Period vs. B-V   & 0.66  & 98.11 & 15 \\
\hline
Cold       & Period vs. Q     & -0.80  & 94.97 & 7 \\
\hline
Centaurs & Period vs. B-V& 0.49 & 91.40 & 14 \\
         & Period vs. e  & 0.47 & 94.04 & 18 \\
\hline
Plutinos & Period vs. H  & -0.48 & 94.14 & 17 \\
\hline
SDO's     & Period vs. V-R  & -0.65 & 91.55 & 8 \\
          & Ampl. vs. H     & 0.57  & 92.86 & 14 \\
\noalign{\smallskip} \hline
\end{tabular}
\end{table}

\begin{table*}
\caption{Some results using the simple model. Listed are the minimum and
maximum rotational period that a Jacobi ellipsoid can rotate. The percentage
of different kinds of objects (Jacobi/MacLaurin/non-equilibrium) are
presented. } \label{model} \centering
\begin{tabular}{lcrccrcr}
\hline\hline \noalign{\smallskip}
Density      & P [min] & P [max] & $\Delta$P & P [eq] & \% Jacobi & \% MacLaurin & \% No Eq.  \\
kg/m$^3$ & hrs & hrs& hrs & hrs & & & \\
 \hline \noalign{\smallskip}
400          & 9.99 & 10.91 & 0.92 & 8.88 &  6.51 & 22.04 & 71.43\\
500          & 8.88 &  9.99 & 1.11 & 7.99 &  9.48 & 26.62 & 63.90 \\
600          & 8.27 &  8.88 & 0.61 & 7.27 &  7.53 & 36.34 & 56.12 \\
700          & 7.50 &  8.27 & 0.77 & 6.66 & 10.33 & 41.29 & 48.37 \\
800          & 7.05 &  7.74 & 0.69 & 6.31 & 10.38 & 46.26 & 43.35\\
900          & 6.66 &  7.27 & 0.61 & 5.85 & 10.28 & 53.52 & 36.19\\
1000         & 6.31 &  7.05 & 0.74 & 5.58 & 12.61 & 55.63 & 31.75 \\
1200         & 5.71 &  6.48 & 0.77 & 5.10 & 14.09 & 62.11 & 23.79 \\
1500         & 5.10 &  5.71 & 0.61 & 4.61 & 11.92 & 72.31 & 15.76 \\
2000         & 4.44 &  5.00 & 0.56 & 4.00 & 10.04 & 82.77 &  7.19 \\
2500         & 4.00 &  4.44 & 0.44 & 3.58 &  6.72 & 90.19 &  3.08\\
3000         & 3.63 &  4.06 & 0.43 & 3.24 &  4.92 & 94.06 &  1.02\\
\noalign{\smallskip} \hline
\end{tabular}
\end{table*}

% -------------------------

\end{document}